\documentstyle[12pt,aasms4]{article}
 
\def\reference{\par\noindent\hangindent=1cm\hangafter=1}

\newcommand{\eq}{\begin{equation}}
\newcommand{\ee}{\end{equation}}

\def\t0{\theta_{\circ}}

\def\be{\begin{equation}}
\def\en{\end{equation}}

\def\gapp{\ \lower 3pt\hbox{${\buildrel > \over \sim}$}\ }
\def\lapp{\ \lower 3pt\hbox{${\buildrel < \over \sim}$}\ }
\lefthead{Jiang}
\righthead{Extrasolar Planets}
 
\begin{document}
 
\title{On the Fate of Close-in Extrasolar Planets}
 
\author{Ing-Guey Jiang$^{1}$, Wing-Huen Ip$^{1}$ and Li-Chin Yeh$^{2}$}

\affil{{$^{1}$ Institute of Astronomy,}\\
{ National Central University, Chung-Li, 
Taiwan} \\  
\ \\
{$^{2}$ Department of Mathematics,}\\
{ National Hsinchu Teachers College, Hsin-Chu, Taiwan}}
 
\authoremail{jiang@astro.ncu.edu.tw}
 
\begin{abstract}


It has been shown that there is a possible mass-period correlation
for extrasolar planets from the current observational data
and this correlation is, in fact, 
related to the absence of massive close-in planets,
which are strongly influenced by the tidal interaction with the central star.
We confirm that the model in P\"atzold \& Rauer (2002) is a good approximation
for the 
explanation of the absence of massive close-in planets.
We thus further determine the minimum possible semimajor axis for these planets
to be detected during their lifetime and also study their migration
time scale at different semimajor axes by the calculations of 
tidal interaction. 
We conclude that the mass-period correlation
at the time when these planets were just formed was less
tight than it is now observed if these orbital migrations are taken into
account. 

 
\end{abstract}
 

\section{Introduction}
 
The number of discovered extrasolar planets is increasing quickly during
recent years. According to the Extrasolar Planets Catalog maintained 
by Jean Schneider (http://cfa-www.harvard.edu/planets/catalog.html), 
in May 2002,
there are about 77 extrasolar planets around 69 main sequence stars.
These planets with mass range from 0.16  to 
17 Jupiter masses ($M_J$) 
 have semimajor axes from 0.04 AU to 4.5 AU and also a wide range
of eccentricities. Interestingly, there is a planet moving on 
an extremely elongated orbit ($e=0.927$) around the solar-type star
HD 80606 (Naef et al. 2001). 
These exciting discoveries provide great opportunities 
to understand the formation and evolution of planetary systems. 

For example, Jiang \& Ip (2001) showed that the interaction with disc is 
important to explain the 
original orbital elements during the planetary formation.
Yeh \& Jiang (2001) analytically showed that 
the scattered planets should in general move on an eccentric orbit and 
thus the orbital circularization must be important for scattered planets
if they are now moving on nearly circular orbits (See Jiang \& Yeh 2002a, 
Jiang \& Yeh 2002b for the following up).



In addition to the dynamical studies, 
Tabachnik \& Tremaine (2002) used the maximum likelihood method
to estimate the mass and period distributions
of extrasolar planets and found there is a mass-period correlation, but they  
attributed their finding to the observational selection effect.

However, Zucker \& Mazeh (2002) claimed that this mass-period correlation
cannot be completely explained by the observational selection effect.
They did some Monte Carlo simulations and show the real dependency between
the mass and period of extrasolar planets. This 
mass-period correlation gives the paucity of massive close-in planets. 
Since they are supposed to be the easiest to detect,  Zucker \& Mazeh (2002)
said this paucity was unlikely to be the result of any selection effect.

P\"atzold \& Rauer (2002) have reported the possible explanation about 
the absence of massive close-in planets by tidal interaction. They 
defined ``critical mass'' to be the maximum mass that the planet can have
and survive under the tidal interaction from the central star
for a given  particular semimajor axis. 
They determined the 
critical mass as function of semimajor axis for some assumed 
stellar dissipation factors and the ages of the planetary systems. 
Their results showed that most planetary systems are located at
the permitted region of the ``critical mass-semimajor axis'' plot 
(their Figure 3) except the $\tau$ Boo system, which needs 
more careful treatment for the assumed parameter values.

However, if these planets could be formed a bit farther from the 
central star initially, they should still survive under the tidal interaction
and thus might be detected during the inward migration.
One should keep in mind that the location where the planets are 
detected are not where they are formed. 
The planets from farther place could migrate inward 
to the region closer to the central star and probably have chances to be 
detected by us.

To further investigate this problem, we carefully
study the planetary 
migration due to tidal interaction. We try to include the effect of 
orbital eccentricity at the beginning and we confirm that 
that the model used in P\"atzold \& Rauer (2002) is a good approximation.
We thus use the similar model in  P\"atzold \& Rauer (2002) for the rest 
calculations. We describe our basic models for tidal interaction in 
Section 2 and the results will be in Section 3. We provide concluding remarks
in Section 4.
 
\section{The Models for Tidal Interaction}

A tide is raised on the central star by the close-in planet because the 
force experienced by the side of the central star facing the 
planet is stronger than that
experienced by the far side of the central star. We consider below the models
for planets on both circular and eccentric orbits.

\subsection{Circular Orbits}

If the close-in planet is 
moving on a circular, equatorial orbit, according to 
the tidal potential theory,  this planet would change its orbit
following below formula:  

\be
\frac{d a}{d t} = 
{\rm sign}(\Omega - n) \frac{3 k}{Q} \frac{m}{M}(\frac{R}{a})^5 n a,
\en
where  $a$ is the semimajor axis, $t$ is the time, $\Omega$ is the 
rotating angular speed of the central star, $k$ is the stellar Love number,
$Q$ is the tidal dissipation function, $m$ is the planetary mass, 
$M$ is the mass of the central star, $R$ is the central star's radius
and $n$ is the orbital mean motion which is determined by
\be
n=\sqrt{ \frac{G(M+m)}{a^3} }.
\en 

We set $k=0.2$ (Murray \& Dermott 1999) and take $Q= 3.0 \times 10^5$
(the average value in  P\"atzold \& Rauer 2002). 

 
The above formula provides a good simple tool to study the tidal orbital decay
for close-in planets. However, in fact, most discovered planets have certain
amount of orbital eccentricities. 
Some of these eccentricities are even very
big. We plan to include the effect  of eccentricity into the 
calculations by the following equations.

\subsection{Eccentric Orbits}
         
We know that the angular momentum is related to orbital eccentricity 
$e$. Thus, the evolution of semimajor axis $a$ due to tidal interaction
should depend on eccentricity $e$ because the tidal torque change the
orbital angular momentum of planets.


The mechanical energy decreasing rate $dE/dt$ due to tidal interaction is 
\be
\frac{d E}{d t} = \Gamma ( \Omega - \frac{d\theta}{dt} ), 
\en
where $\Gamma$ is the magnitude of the torque, 
$\Omega$ is the spin angular speed of the central star and 
$d\theta/dt$ is the orbital angular speed of the planet at particular time.

$\Gamma$ can be approximated by:
\be
\Gamma = \frac{3}{2}k \frac{Gm^2}{a^6}R^5\frac{1}{Q},
\en
these parameters are defined in last sub-section.

The orbital angular speed of the planet can be expressed as  
\be
\frac{d\theta}{dt} = \frac{h}{r^2},
\en
where $h=\sqrt{ G(M+m) a (1-e^2)}$  and $r$ is approximated as 
$r=a(1-e\cos nt)$. 

Therefore, 
\be
\frac{d E}{d t} = \frac{3}{2}k \frac{Gm^2}{a^6}R^5\frac{1}{Q}
[ \Omega - \frac{\sqrt{ G(M+m) a (1-e^2)}}{  a^2(1 - e\cos nt)^2} ]
\en 

On the other hand, the mechanical energy of the system can be expressed 
as 

\be
E= \frac{1}{2} I \Omega^2 - G\frac{Mm}{2a} 
\en
and

\be
\frac{dE}{dt} = I \Omega \frac{d\Omega}{dt} + G\frac{Mm}{2a^2} \frac{d a}{dt}
\en
By Kepler's third law, 
\be
G(M+m) = n^2 a^3,
\en
we have 
\be
\frac{dE}{dt} = I \Omega \frac{d\Omega}{dt} + 
\frac{Mm}{2(M+m)} n^2 a \frac{da}{dt}.
\en

Further, the angular momentum of the system is
\be
L= I\Omega + \frac{Mm}{M+m} a^2 n (1-e^2)^{1/2}, 
\en
where $I$ is the moment of inertia of the central star, $e$ is the orbital 
eccentricity
and we have ignore the contribution from the spin of the planet.

By the conservation of angular momentum, $dL/dt=0$, we have

\be
I\frac{d \Omega}{d t} = - \frac{1}{2}
\frac{Mm}{M+m}na\frac{d a}{d t}\sqrt{1-e^2}  
         + \frac{Mm}{M+m}n a^2\frac{e de/dt}{\sqrt{1-e^2}}. 
\en

In general, both terms on the right hand side of Equation (12) should 
be considered. 
The second term divided by the first term would be 
\be
e^2(1-e^2)[ \frac{63}{6}\frac{Q}{k \mu_p Q_p}(\frac{M}{m})^2
(\frac{R_p}{R})^5 ],  
\en
where Equation (4.198) in Murray \& Dermott (1999) has been used to estimate
the value of $de/dt$ and we use $\mu_p$, $Q_p$ and $R_p$ etc. to replace 
the corresponding parameters
$\tilde{\mu_s}$, $Q_s$ and $C_s$ etc.
of Equation (4.198) in Murray \& Dermott (1999).
If we use the Jupiter as an example, this ratio would be about 1 when 
$e=0.1$ and $Q/(k\mu_p Q_p)=1$.
  
We plan to consider the simple case when $e^2 Q/(k\mu_p Q_p)$ 
is small enough and the second 
term can be ignored. We will leave more general case in which both orbital 
migration and circularization need to be included to the future work.


Thus,
\be
\frac{dE}{dt} = \frac{1}{2}( n - \Omega\sqrt{1-e^2} ) 
                \frac{Mm}{M+m}na \frac{d a}{dt}
\en

From Equation (14) and Equation (6), we have 

\be
\frac{d a}{d t} = 
3k \frac{Gm^2}{a^7}\frac{R^5}{Qn} \frac{M+m}{Mm}
(n-\Omega\sqrt{1-e^2})^{-1} [\Omega - 
 \frac{\sqrt{ G(M+m) a (1-e^2)}}{ a^2 (1 - e\cos nt)^2}],
\en
where
$\Omega$ is related to $a$ by Equation (11). 



Given an assumed initial angular momentum $L$ etc.,
$a$ can be solved numerically  by Equation (15). 

\section{Results}
 
By the equations in last section, we can study the inward migration of planets
due to tidal interaction. We place the planet at 
different initial semimajor axis as different case: 
0.02, 0.03, 0.04, 0.05 and also 0.06 AU. 
Figure 1 are the plots of semimajor axis as function of time
for these different initial semimajor axes when we set the planetary mass to 
be particular value.
Thus, there are five curves on each panel of Figure 1.
Figure 1(a)-(d) are the results
when the planetary masses are assumed to be 5$M_{J}$, 2$M_{J}$, 
$M_{J}$,  0.5$M_{J}$ individually. 
Since 2 Gyrs is about the 
age of $\tau$ Boo system and thus we regard 2 Gyrs as the 
typical age of extrasolar planetary systems. Those planets who can survive
for 2 Gyrs under the tidal interaction are possible to be detected.

All the curves are the results when we assume the planets move on circular 
orbits and the triangle points are the results when 
the planets move on eccentric orbits (assume $e=0.5$ and $e^2 Q/(k\mu_p Q_p)$
is small enough).
In general, the results of eccentric orbits
 are quite similar to the results of 
 circular orbits and the ignorance of eccentricity will not affect the 
determination of planet surviving time scale etc. 
This confirms that the equations used in 
P\"atzold and Rauer (2002) are good approximations and we thus 
use the model of circular orbits for all the rest calculations. 
 
Figure 1(a)-(d) show that when initial semimajor axis $a_i=0.06$ AU the planet
would only have tiny migration during 2 Gyrs. The planet can easily survive
under the tidal interaction. If 
initial semimajor axis $a_i=0.05$ AU, the orbital semimajor axis
decays a bit more.
If the initial semimajor 
axis $a_i=0.04$ AU, the planet fall into the central star when $t$ is about 
1 Gyrs for the case of 5$M_{J}$ but still survive for all other cases.
When  initial semimajor axis $a_i=0.03$ AU, 
the planet falls into the central star
within 1.5 Gyrs.  If initial semimajor axis $a_i=0.02$ AU, the planet almost
approaches to the central star immediately. 

The detection probability for particular range of semimajor axis 
depends on how much time the planet can survive around that range.
We plot the time the planet should spend from one semimajor axis $a_j$ to
another semimajor axis $a_{j+1}$ (we assume $a_j > a_{j+1}$) during the 
orbital decay in Figure 2. There are two sets of $a_j$: one makes 
$\delta a\equiv a_j-a_{j+1} = 0.005$ AU (dotted lines), another set 
$\delta a\equiv a_j-a_{j+1} = 0.0025$ AU (solid lines). 
Figure 2(a)-(d) are  
the results when we set the planetary mass to be 5$M_{J}$, 2$M_{J}$,
$M_J$, 0.5$M_J$ individually.

In Figure 2(a)-(b), the planet spends more than 1 Gyrs to stay around 0.05 AU
and thus the planet is likely to survive during 2 Gyrs. 
However, the planet only stays around $0.04$ AU for about 0.5 Gyrs
and around $0.03$ AU for about 0.2 Gyrs only. These time scales are 
considerablely smaller than the age of the planetary system and thus
the planet initially formed around these locations are 
very unlikely to be observed. 

When the planetary mass is smaller as in Figure (c)-(d), 
the planet can survive for much longer (more than 1.5 Gyrs) around 0.04 AU
and still only stays around 0.03 AU for order of 0.5 Gyrs. 
This implies that the probability that the planet is detected to be around
0.03 AU is very small.

Figure 3 are the ${\rm ln}(a/{\rm AU})-{\rm ln}(M/M_J)$ 
plots for all discovered 
extrasolar planets, where $M$ is the planetary mass and $a$ is the 
semimajor axis. 
The data for these planets are from 
Extrasolar Planets Catalog 
(http://cfa-www.harvard.edu/planets/catalog.html) in May 2002.
 In Figure 3(a), we take $a$ to be the values of 
current semimajor axes of these discovered extrasolar planets. 
However, in Figure 3(b)-(d), 
we take
$a$ to be the planetary semimajor axes backward in time for 2, 6, 12 Gyrs
individually.
The values of $a$ backward in time can be obtained by
Equation (1).  

In Figure 3(b)-(d), we found that 
most planets do not move on the ${\rm ln}(a/{\rm AU})-{\rm ln}(M/M_J)$
 plane but some of them do 
move a lot when they are backward in time. 

It is quite obvious that the planets line up on the left side of the 
plots in Figure 3(b)-(d) and the position of this line hardly moves
from Figure 3(b) to 3(d). This line can be approximated by
\be
{\rm ln}(M/M_J) = \frac{1}{5}[ 28\  {\rm ln}(a/{\rm AU}) + 62 ].
\en
We can also see that those planets do not move
much are all on the right side of this line. 
This line can thus be regarded as the ``critical line'': all planets on 
the right side of this line would not migrate much during their lifetime
but all the planets standing on this critical line of Figure 3(b)-(d)
would move to the left-up corner of Figure 3(a) after 2, 6 or 12 Gyrs
and finally all the planets on the
left side of this line would migrate inward quickly to approach the central 
star and thus cannot be detected.

 

\section{Concluding Remarks}

As dynamical friction successfully explained the orbit of 
Sagittarius dwarf galaxy (Jiang \& Binney 2001), 
the tidal interaction can indeed explain the 
current observed mass-period correlation reported by Zucker \& Mazeh (2002).
The results in Figure 1 give us the full picture of inward migration 
due to tidal interaction. We found that 0.03 AU seems to be the critical
semimajor axis for the planet with mass of order of $\tau$ Boo system 
to survive
in 2 Gyrs. This is consistent with the current observational results that the
smallest semimajor axis of discovered planet is about 0.04 AU.

On the other hand, we can also check this minimum possible semimajor axis
from another point of view. In Figure 2, 
the time scale for a planet can survive is smaller if the planet is 
closer to the central star initially
and  the time a planet
can stay around 0.03 AU is considerablely much less than 2 Gyrs,
which was regarded as the typical age of these planetary systems.  
Because time scale is too short, the probability to detect the 
planet is very small. 

Moreover, we interestingly discover the observational ``critical line'' 
on  ${\rm ln}(a/{\rm AU})-{\rm ln}(M/M_J)$  plane. 
All the planets on the left side of this line would migrate inward quickly 
to approach the central star and thus cannot be detected.

Therefore, the initial configuration on ${\rm ln}(a/{\rm AU})-{\rm ln}(M/M_J)$
plane might be composed of 
all the points on Figure 3(b) plus those points which might have been 
on the left side of the ``critical line'' about 2 Gyrs ago but disappear
in Figure 3(a) because these planets already fall into the central star.
From this point of view, even there is correlation between mass and period 
for current discovered planets as claimed by Zucker \& Mazeh (2002),
this  correlation could be weaker or less obvious at the time 
when these planets
were just formed since we can add arbitrary number of ``possible'' planets
on the left side of our observational ``critical line'' if there is no
difficulty to form planets there in theory. 
This tells us that we should be careful when we try to link the
mass-period correlation to the theory of planetary formation. 

\section*{Acknowledgment}
We are grateful to the referee's good suggestions.
This work is supported in part 
by the National Science Council, Taiwan, under Grants NSC 90-2112-M-008-052.

 
\clearpage
\section*{REFERENCES}

\begin{reference} 
\reference Jiang, I.-G. \& Binney, J., 2000, MNRAS, 314, 468  
\reference Jiang, I.-G. \& Ip, W.-H., 2001, A\&A, 367, 943 
\reference Jiang, I.-G. \& Yeh, L.-C., 2002a, 
Int. J. Bifurcation and Chaos, in press,  astro-ph/0201022
\reference Jiang, I.-G. \& Yeh, L.-C., 2002b, in preparation
\reference Murray, C. D. \& Dermott, S. F., 1999, Solar System Dynamics
(Cambridge: Cambridge University Press)
\reference Naef, D. et al., 2001, A\&A, 375, L27
\reference P\"atzold, M. \& Rauer, H., 2002, ApJ, 568, L117 
\reference Tabachnik, S. \& 
Tremaine, S., 2001, submitted to AJ, astro-ph/0107482
\reference Yeh, L.-C. \& Jiang, I.-G., 2001, ApJ, 561, 364
\reference Zucker, S. \& Mazeh, T., 2002, ApJ, 568, L113

\end{reference}

\clearpage
\begin{figure}[tbhp]
\epsfysize 7.0in \epsffile{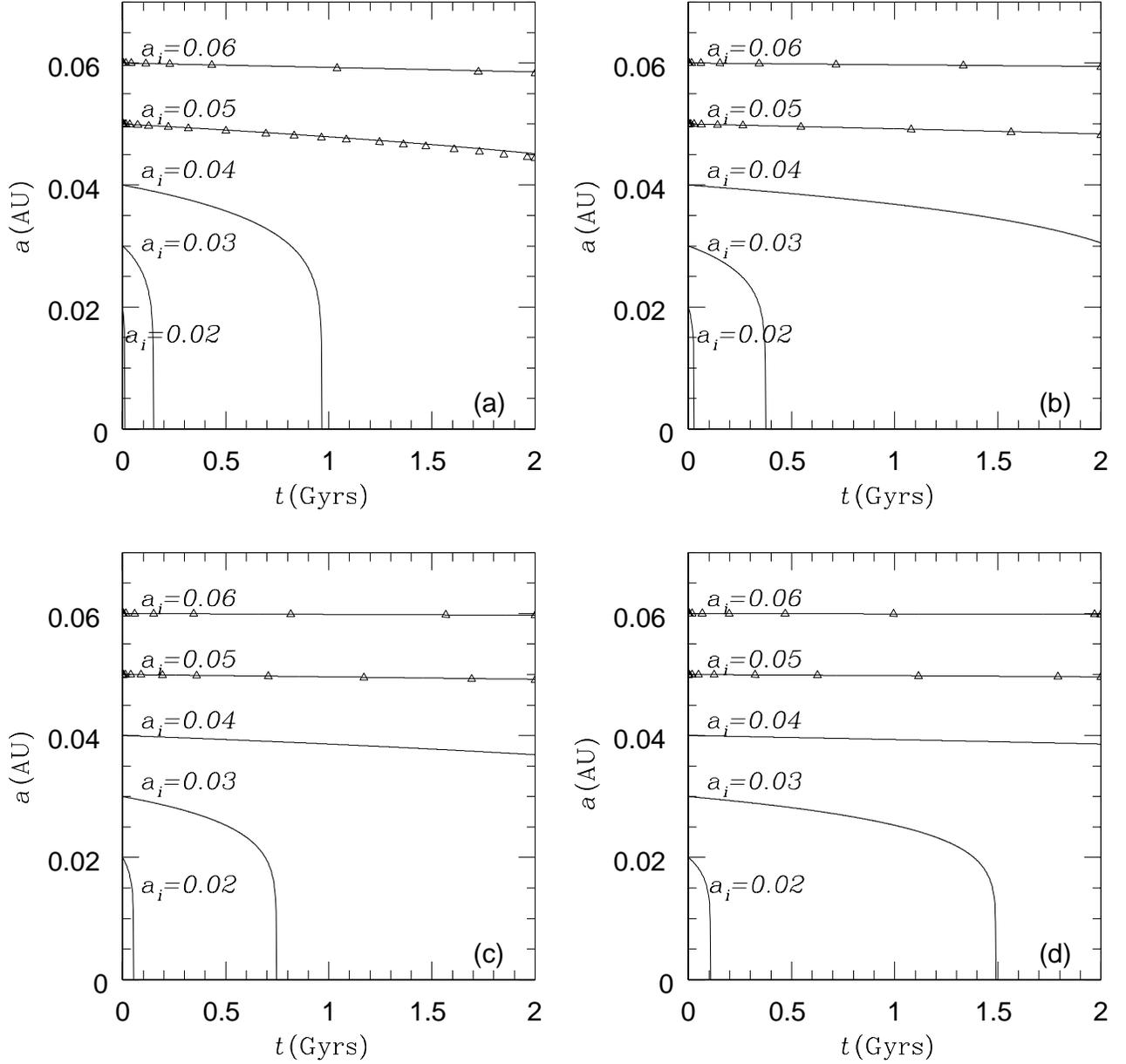}
\caption{The semimajor axis as function of time. The solid curves
are the results of $e=0$ and initial semimajor axis 
$a_i=0.02, 0.03, ..., 0.06$. The triangle points are the results of $e=0.5$
and $a_i=0.05, 0.06$.
(a) Planetary mass is 5$M_J$,
(b) Planetary mass is 2$M_{J}$,
(c) Planetary mass is $M_J$,
(d) Planetary mass is 0.5$M_J$. 
 }
\end{figure}
 
\clearpage
\begin{figure}[tbhp]
\epsfysize 7.0in \epsffile{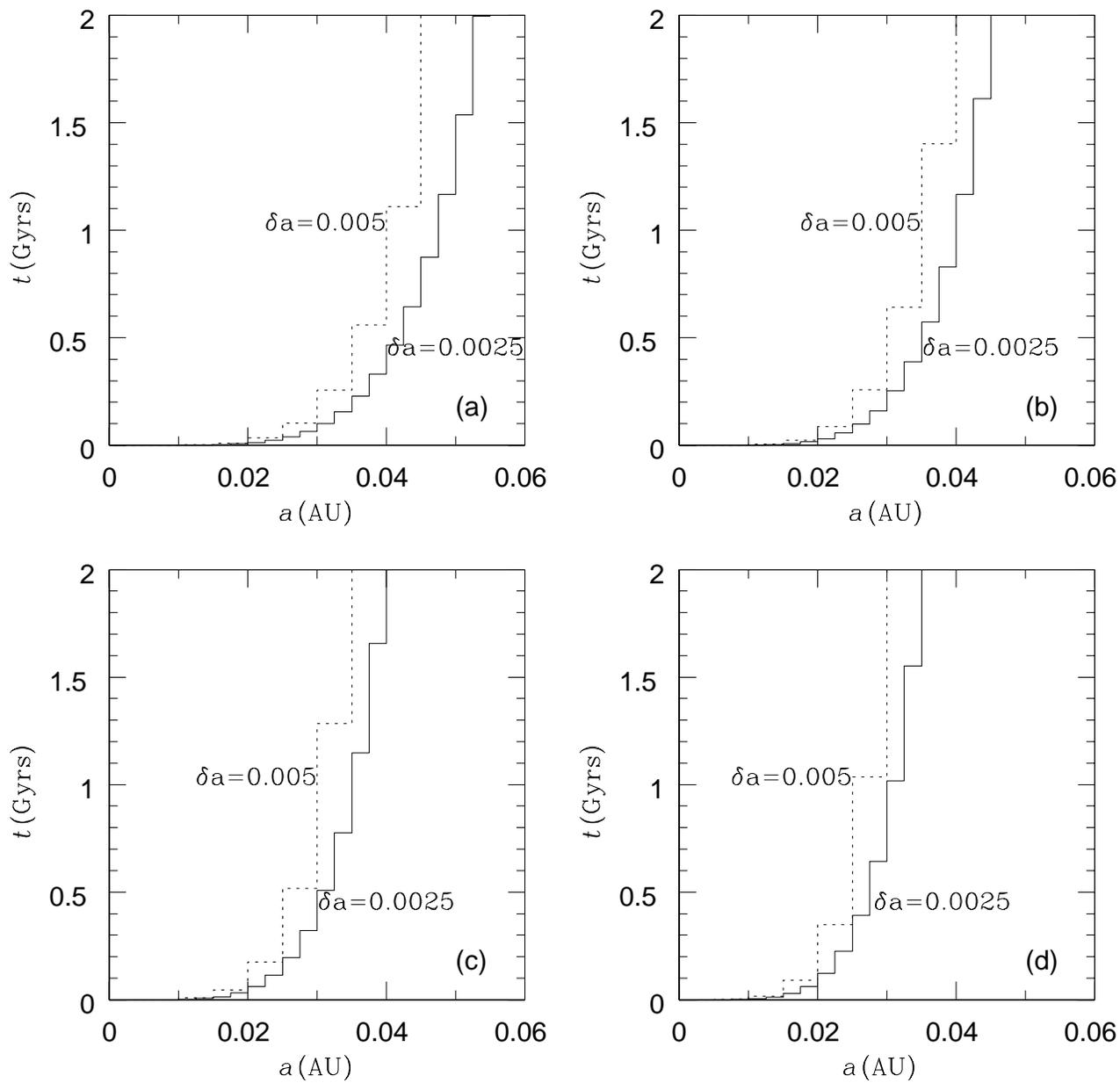}
\caption{The migration time scales for give ranges of semimajor axis.
The dotted lines are the results when $\delta a= 0.005$ and 
the solid lines are the results when $\delta a= 0.0025$.
(a) Planetary mass is 5$M_J$,
(b) Planetary mass is 2$M_{J}$,
(c) Planetary mass is $M_J$,
(d) Planetary mass is 0.5$M_J$. 
}
\end{figure}

\clearpage
\begin{figure}[tbhp]
\epsfysize 7.0in \epsffile{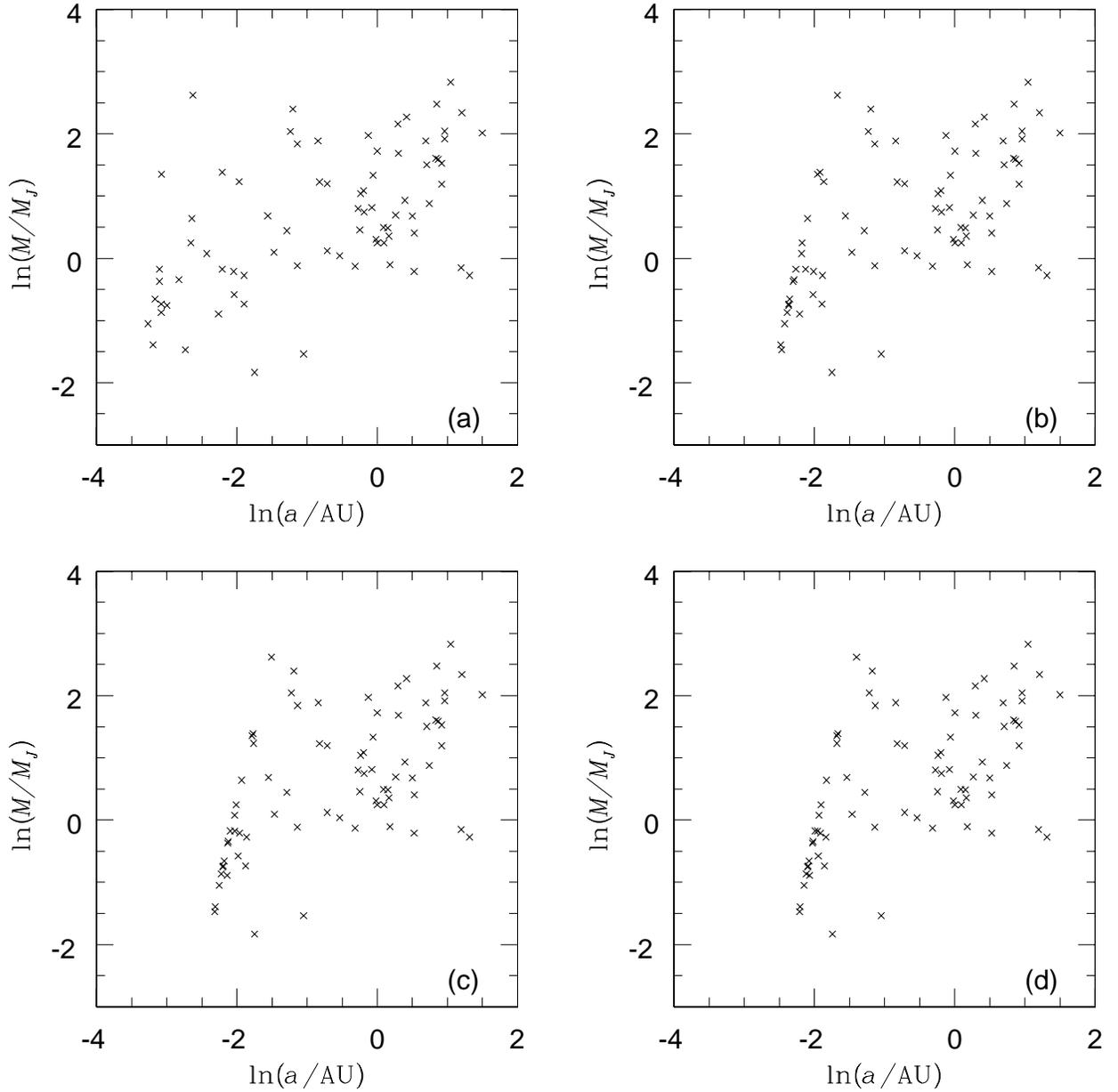}
\caption{The 
${\rm ln}(a/{\rm AU})-{\rm ln}(M/M_J)$ plot for all discovered extrasolar 
planets in Extarsolar Planets Catalog in May 2002.
(a) Current discovered configuration,
(b) Backward in time for 2 Gyrs,
(c) Backward in time for 6 Gyrs,
(d) Backward in time for 12 Gyrs.
}
\end{figure}

\end{document}